\documentstyle[11pt,newpasp,twoside, psfig]{article}
\def\edcomment#1{\iffalse\marginpar{\raggedright\sl#1\/}\else\relax\fi}
\reversemarginpar
 
\begin{document} 
\title{Studying the Nearby Universe with {\sl Chandra}}
 \author{Q. Daniel Wang}
\affil{Astronomy Department, University of Massachusetts, Amherst, MA 01003}

\begin{abstract}
I highlight results from {\sl Chandra}
observations of nearby galaxies, including the Milky Way. These observations
have offered insights into old mysteries and indications of new high energy 
astrophysical phenomena and processes that are yet to be understood. 
\end{abstract}

\section{Introduction}

X-ray emission is an excellent tracer for high temperature and/or high velocity
activity in galaxies. Such high energy activity is largely produced by the 
end-products of massive stars and by supermassive 
black holes in galactic centers (e.g., Helfand 1984). Neutron
stars, stellar mass black holes, supernova remnants (SNRs), and the 
diffuse hot interstellar medium (ISM), as well as active nuclei (AGNs) 
are often best studied in X-ray.
It is not unusual that such objects or gaseous features leave no trace
at all in other wavelength bands. The study of massive star end-products, in particular, is important to the understanding of the interplay between stars
and the ISM, or the ``ecosystem'' of galaxies. This is also intimately tied to 
the thermal and chemical feedback from galaxies to the intergalactic medium
(IGM).
Significant X-ray emission may also arise from many other processes such as
magnetic field reconnection and the bremsstrahlung of low energy cosmic rays. 

{\sl Chandra} is a unique X-ray observatory that for the first time
provides us with the capability to obtain X-ray images with arcsecond 
resolution, broad energy coverage (0.2-10 keV), and good CCD spectral resolution 
($\sim 130$ eV). The data obtained with  {\sl Chandra} is revolutionizing our 
view of the X-ray universe. In this writing-up of my overview talk,
I will concentrate on various new high energy astrophysical phenomena 
revealed by {\sl Chandra} ACIS observations, particularly on topics that 
are not covered (adequately) by other speakers. I will start with our 
Galactic center region and then move on to nearby galaxies, making
connections to the evolving universe that Professor Hasinger will be reviewing
next.

\section{The Milky Way Center as a Laboratory for High Energy Astrophysics}

Nuclear regions of galaxies are the mecca of high energy phenomena and
processes, which are  manifested by AGNs and starbursts. 
The best site for a detailed study of the complex 
interaction of such energetic activity and its interaction with 
the unique nuclear environment is
our own Galactic center (GC) region at a distance of only 8 kpc. 

An overall picture of the GC region as a mildly active starburst galaxy 
has emerged. Three young massive stellar clusters have been discovered
 (Arches, Galactic center, and Quintuplet; e.g., Figer et al. 1999).
Massive stars ($M \ga 10 M_\odot$) themselves are energetic objects, 
releasing fast stellar winds and large amounts of ionizing photons. 
Such stars are also short-lived and
explode as supernovae on time-scales from a few to a few tens of million
years, depending of their masses. 
Stellar remnants of the explosions (neutron stars and
black holes) can often be seen as pulsars and 
X-ray binaries, whereas supernova blastwaves can naturally heat 
ambient gas to temperatures of million degrees or higher. Radio and 
infrared observations have already revealed luminous shells or arcs as 
well as nonthermal filaments (NTFs) with magnetic field 
strengths possibly as high as $\sim 1$ mG (e.g., Morris 1994). These 
features, uniquely seen in the GC region, demonstrate 
the strong interaction between massive stars and the extreme
high density and high magnetic field environment. 
{\sl Chandra} has provided us with a new tool to probe 
the high-energy properties of the GC. 

Wang, Gotthelf, \& Lang (2002) have carried out a systematic survey of a 
2 $\times$ 0.8 square degree field, which consists of 30 overlapping 
{\sl Chandra} 
ACIS-I observations (about 11 -- 12 ks exposure each) and covers the most 
active ridge around the GC (Fig.\ 1). The survey, complemented by 
deeper pointings ($\ga 50$ ks each) on several specific targets 
with the same instrument (Yusef-Zadeh et al. 2002; Takagi, Murakami, 
\& Koyama 2002; Baganoff et al. 2001), provides us with
the first high resolution X-ray view of the GC with sensitivity 
more than two orders of magnitude higher than any pre-{\sl Chandra} observations 
(e.g., {\sl ASCA}). While the {\sl Chandra} data analysis is still ongoing, 
I summarize some of the preliminary results.

\subsection{A Global X-ray View}

\begin{figure*}[bht]
 \begin{center}
 \end{center}
\caption{
{\sl Chandra} ACIS-I intensity contours overlaid on 
the radio (90 cm) image of the GC ridge. The X-ray intensity
is calculated in the 1--8 keV range and is adaptively smoothed
with a signal-to-noise ratio of $\sim 3$. The saw-shaped boundaries of 
the map, plotted in Galactic coordinates, results from a specific roll 
angle of the observations.
}
\end{figure*}

Fig.\ 1 includes a broad-band X-ray contour image constructed with the
data from the GC ridge survey. Close to $10^3$ discrete sources are detected. 
Only a handful of these X-ray sources are previously known.
The brightest source 1E 1740.7-29.42 is a ``micro quasar'',
which contains an accreting stellar mass black hole and
extended radio jets. The second brightest source 1E 1743.1-2842 is 
a low-mass X-ray binary, possibly with a neutron star as the primary.
SNR G0.9+0.1 is a composite supernova remnant with an X-ray-emitting 
Crab-like nebula and a larger-scale radio shell. 
Several of the other luminous X-ray sources detected previously were 
relatively faint during the {\sl Chandra} survey.
GRO J1744-28 and SAX J1747.0-2853 are two examples 
(Wijnands \& Wang 2002; Wijnands, Miller, \& Wang 2002). 
The 0.5-10 keV luminosity of the $\gamma$-ray source 
GRO J1744-28 during the survey was only $2 - 4 \times 10^{33} {\rm~erg~s^{-1}}$ 
(assuming that the source is physically near the GC),
compared to the earlier {\sl ROSAT} detection of $\sim 2 \times 10^{37} 
{\rm~erg~s^{-1}}$ in the 0.1-2.4 keV band. The {\sl Chandra} survey observation
of SAX J1747.0-2853, with a  0.5-10 keV luminosity of 
$3 \times 10^{35} {\rm~erg~s^{-1}}$, was taken between two outbursts in 
2000 and 2001. Another potentially interesting source is 
CXOGCS J174502.3-285450, which appeared orders of magnitude brighter in 
another {\sl Chandra} observation taken about a year earlier than the survey. 
The ``quiescent'' properties of these transient sources can now be used to 
place interesting constraints on the cooling process of neutron star surfaces and 
on the X-ray emission mechanism of neutron stars in poorly studied state
of low accretion rates. The accurate {\sl Chandra} positions of 
the sources will also make it much easier to identify 
counterparts at other wavelengths.

What are the origins of those newly detected and 
relatively faint X-ray sources? A considerable 
fraction (up to $\sim 30\%$) of the sources are bright in 1-3 keV band
and thus must be foreground objects (e.g., cataclysmic variables and 
normal stars). Many of the soft X-ray bright foreground
stars are associated with two stellar clusters embedded within known optical
emission nebulae Sh2-20 and Sh2-17 (Dutra \& Bica 2000 and references therein;
Wang et al. 2002). Sources that are located physically near the GC
or beyond are subject to X-ray absorption of equivalent total gas column density
greater than a few times $10^{22}
{\rm~H~atoms~cm^{-2}}$ and must show little or no soft X-ray emission/flux. 
A preliminary estimate based on millimeter and 
far-infrared emissions in the GC field suggests that less than a few 
percents of all
these sources were extragalactic in origin. Therefore, 
the bulk of the sources are likely located in the GC region.

Pfahl et al. (2002) proposed that many of the sources may be wind-accreting 
neutron stars. An 
infrared observing campaign is now under way to search for the predicted
stellar counterparts of the hard X-ray sources. The confirmation or rejection 
of a proposal like this is especially important for massive binary population 
synthesis studies. 

\subsection{Sgr A$^*$}

	The single most important discovery made by {\sl Chandra} 
in the GC region is probably the rapid X-ray flaring from Sgr A$^*$,
which hosts the central black hole of $\sim 3 \times 
10^6 {\rm~M_\odot}$ (Baganoff et al. 2001). The flare detected in 2000 lasted 
for about 3 hours and exhibited variability on time scales of
a few minutes. The X-ray emission of the flare, characterized by a power law 
spectral model, has a peak 2-10 keV 
luminosity of $\sim 1 \times 10^{35} {\rm~erg~s^{-1}}$, or about 45 times
the quiescent-state luminosity of Sgr A$^*$.
The detection of such a flare provides the most compelling evidence that 
matter falling toward the black hole is fueling energetic activity. 

However, the nature of the quiescent-state X-ray emission, which is slightly 
extended on a scale of $\sim 1^{\prime\prime}$, remains unclear.
The X-ray spectrum shows evidence for an emission line at $\sim 6.6$ keV and 
a characteristic thermal plasma temperature of $\sim 4$ keV.  
The faintness of the emission,
together with the linear and circular polarization measurements of Sgr A$^*$ in
millimeter and radio wavelengths, places tight constraints on the 
GC black hole accretion rate, $3 - 4$ orders of magnitude less than 
the expected Bondi accretion rate of $\sim 10^{-4} - 10^{-5} 
{\rm~M_\odot~yr^{-1}}$. Furthermore, a considerable fraction of the 
quiescent-state emission should be due to normal stars in the GC stellar 
cluster and possibly to the expected presence of numerous compact stellar 
remnants, e.g., millisecond pulsars around the black hole, as 
in globular clusters (e.g., Grindlay et al. 2001). Such stellar remnants,
if present, may generate powerful relativistic winds that could significantly 
reduce the amount of gas falling into the black hole.

\subsection{Arches Cluster}

{\sl Chandra} has also detected the strong X-ray emission from the Arches 
stellar cluster and its vicinity (Wang et al. 2002; Yusef-Zadeh et al. 2002). 
Located at only 12$^\prime$ away from  Sgr A$^*$, the cluster contains more 
than 100 O-type and Wolf-Rayet stars within a region of $\la 1$ pc across 
(Figer et al. 1999; Fig.\ 2). At an age of $1 - 2 \times 10^6$ years, 
the cluster is responsible for the ionization of the Arched filaments --- 
the most prominent thermal features in the GC region. The Arches cluster is 
thus an excellent testbed for understanding physical processes involved 
in a compact young massive stellar cluster and in its interplay with the 
unique GC environment. 

\begin{figure*}[!hbt]
\unitlength1.0cm
    \begin{picture}(13,6.5) 
\put(0,0){
          \begin{picture}(6.5,6.5)
	\end{picture}
	}
\put(6.5,0){
          \begin{picture}(6.5,6.5)
	\end{picture}
	}
    \end{picture}
\caption{
{\bf Left panel:} The Arches cluster and its vicinity. 
The grey-scale image represents the {\sl Chandra} ACIS-I image 
in the 1-8 keV band, while the overlaid contours illustrate the the 6.4-keV line 
intensity distribution. 
{\bf Right panel:} The close-up of the region outlined by 
the central box in the left panel. The contours are from the 1-8 keV band image
and the grey-scale image from an HST NICMOS near-IR observation (Figer et al. 1999).}
\end{figure*}

Fig.\ 2 shows three distinct X-ray sources/peaks apparently associated with 
the cluster. The spectra of these sources 
exhibit the prominent K$\alpha$ emission 
lines of highly-ionized ions such as Ca XIX and Fe XXV
(Yusef-Zadeh et al. 2002). Each source has an unabsorbed 
X-ray luminosity of $\sim 10^{35} {\rm~ergs~s^{-1}}$ in the 0.2--10 keV band.
There is also a very extended diffuse X-ray component with a 
luminosity of $\sim 2 \times 10^{34} {\rm~ergs~s^{-1}}$. The spectrum
of this component presents a strong 6.4-keV emission line with an equivalent 
width of $\sim 1$ keV (Yusef-Zadeh et al. 2002; see also  Fig.\ 2).

The nature of the X-ray emission from the Arches cluster remains largely uncertain.
A considerable fraction of the emission may come 
directly from individual massive stars. The total luminosity of the
cluster ($L_{bol} \sim 4 \times 10^{41} {\rm~ergs~s^{-1}}$; Figer et al.
1999) and the ``canonical'' relation between X-ray and total luminosity of 
$L_x/L_{bol} \sim 10^{-7}$ (Chlebowski 
1989) predicts an integrated X-ray luminosity 
$L_{x} \sim 4 \times 10^{34} {\rm~ergs~s^{-1}}$, or 
about 10\% of the observed X-ray luminosity of the Arches cluster. 
This fraction could be higher if the relation under-estimates the X-ray 
luminosity of stars with high metal abundances ($\sim 3 \times$ solar)
and in a dense stellar environment (e.g., Chlebowski 1989). 

High X-ray luminosity sources may represent colliding winds in
close binaries of 
extremely massive stars (Portegies Zwart et al. 2001). However, the X-ray 
centroids of the sources in the Arches cluster do not
match the positions of the brightest near-infrared stars (Fig.\ 2), 
which typically have the strongest stellar winds (Lang et al. 2001). 

Alternatively, the X-ray sources may represent diffuse emission peaks of the 
so-called cluster wind (Raga et al. 2001; Yusef-Zadeh
et al. 2002). Because of the high density of the massive stars, their
stellar winds collide with each other and can be partly thermalized to an 
initial temperature of a few times $ 10^7$ K. The expanding of this
hot gas may be considered as a wind from the entire cluster. 
X-ray emission from the cluster wind of the Arches cluster has been 
simulated by Raga et al. (2001). While it is not difficult to tune the
stellar wind velocity and intensity to match the characteristic 
temperature and luminosity inferred from the X-ray data, the spatial
distribution of the simulated X-ray morphology does not seem to match the
observation. For example, the observed bright northern 
source, clearly resolved, is totally absent in all three simulations with 
assumed different line-of-sight distributions of cluster stars. The simulated 
X-ray emission also appears considerably less extended than
the observed, linearly by a factor of $\sim 2$.
Furthermore, the strong 6.4-keV fluorescence line from cold Fe atoms in 
the spectrum of the diffuse X-ray emission is
inconsistent with the hot thermal plasma model of the cluster wind.

The 6.4-keV line and  much of the diffuse emission are most likely
due to the X-ray fluorescence/scattering of the cluster X-ray emission 
by dense and cold gas in the vicinity. Indeed, the emission seems to 
 coincides spatially with the southern tip of a dense molecular cloud observed 
in the region (Serabyn \& G\"usten 1987). An ongoing in-depth study with 
recently obtained high 
resolution data on the cloud will allow for a detailed modeling 
of the fluorescence/scattering processes. 

\subsection{X-ray Threads}

{\sl Chandra} observations have revealed a number of very interesting 
linear X-ray features, which we call X-ray threads. They 
are oriented more-or-less vertically with respect to the Galactic plane, 
in a fashion similar to the radio bright NTFs.
Radio polarization measurements show that the NTFs represent 
synchrotron radiation from relativistic particles trapped in magnetic field
flux tubes. But the origins of both the magnetic field and the particles 
are still a mystery. 

What might be the relationship 
between the radio NTFs and the X-ray threads?
Interestingly, the three brightest X-ray threads 
(G0.13-0.11, G359.89-0.08, \& G359.55+0.17)
all seem to be related to either NTFs or nonthermal radio ``wisps''. 
The X-ray spectra of these threads can be characterized by a simple 
power law, consistent with being nonthermal in nature. 
Wang, Lu, \& Lang (2002) have proposed that G0.13-0.11 may represent the 
leading-edge of a pulsar wind nebula,
The putative pulsar, identified as a point-like X-ray source embedded 
in the thread, is probably moving in a strong magnetic field environment.
The main body of this pulsar wind nebula is likely traced 
by a bow-shaped radio feature, which is apparently bordered by
G0.13-0.11 and is possibly associated with the prominent NTFs of the Radio 
Arc. The high energy pulsar wind and the subsequent 
reverse-shock particle acceleration provide a natural 
explanation for the ultra-relativistic particles ($\sim 10^{14}$ eV) required
for producing synchrotron X-rays.

A similar interpretation may also apply to G359.55+0.17 (Lu, Wang, \& Lang
2002). This X-ray thread lies between an elongated nonthermal radio 
``wisp'' and a possible point-like X-ray source (CXOGCS J174539.7-290413). 
In this case, the pulsar may be moving in a direction nearly parallel 
to magnetic field lines in the region. The ram-pressure confinement of the
shocked pulsar wind material can then produce the linear structure formed by
the X-ray source, the X-ray thread, and the radio wisp. The offset between
the radio and X-ray features, for example, can be naturally explained by the 
synchrotron lifetime difference of the corresponding electrons/positrons.

Even more intriguing is the X-ray thread G359.54+0.18, which 
coincides exactly with the brightest section of one of the two adjacent 
and parallel NTFs (Wang 2002). While inverse self-Compton scattering is found 
unimportant, the X-ray emission may also 
be synchrotron in nature. However, it is not clear what might be the origin of 
the required ultra-relativistic particles. A recent 
magnetic field reconnection between the two NTFs, as indicated by their 
close proximity, may be a possibility. 

\subsection{Diffuse X-ray Emission}

\begin{figure*}[!thb]
 \begin{center}
 \end{center}
\caption{
Narrow-band {\sl Chandra} ACIS-I intensity maps in the 2.36-2.56 keV,  
6.2-6.55 keV, and  6.55-6.9 keV ranges (upper, middle, and lower panels), 
which are dominated by emission lines of S XV K$_\alpha$, cold Fe K$_\alpha$, 
and Fe XXV K$_\alpha$, respectively. 
The rest is the same as in Fig. 1. 
}
\end{figure*}

The total count contribution from detected discrete sources from the GC region,
excluding the two brightest ones (1E 1740.7-29.42 and 1E 1743.1-2842), 
is only about 10\% of the remaining X-ray flux, which has a total 2-10 keV 
luminosity of $\sim 10^{37} {\rm~ergs~s^{-1}}$. 
What might be the origins of this apparently diffuse component?
 
\begin{itemize}
\item The component cannot be explained by any known populations of X-ray sources 
with individual fluxes below our detection limit ($L_{x} \sim 10^{32-33} 
{\rm~ergs~s^{-1}}$).

\item The spectrum of the component shows strong
S XV K$_\alpha$ and Ar XVII K$_\alpha$ lines, relative to the
Fe XXV K$_\alpha$ line (Wang 2002; e.g., Fig.\ 3). This indicates 
that the diffuse soft ($\la 4$ keV) X-ray emission is dominated by 
thermal plasma at temperatures $\la  6 \times 10^7$ K. This plasma, however, 
cannot account for all the continuum emission at higher energies. 

\item The lack of a detailed correlation of the X-ray emission with either 
radio continuum (except for G359.54+0.18; see \S 2.4) or near-infrared features 
further suggests that inverse Compton scattering is not important. 

\item The required total X-ray luminosity for producing the 6.4-keV line 
emission (e.g., Fig.\ 3) via fluorescence is about two orders of
magnitude greater than what is observed. Of course, some X-ray transients could 
be very luminous in the past. Alternatively, Sgr A$^*$ could have
a luminosity of several times $10^{39} {\rm~ergs~s^{-1}}$ about 300 years ago, or a factor of $\sim 10^6$ brighter than the average luminosity
at present (Koyama et al. 1996). However, this scenario would predict 
an intensity correlation of the 6.4-keV line emission with molecular 
tracers. This is not apparent at least in parts of the GC region (Wang 2002).

\item The 6.4-keV line emission may arise from the filling of 
K-shell vacancies generated by collisions between cold iron atoms
and low energy cosmic rays (Valinia et al. 2000). The bremsstrahlung of
these cosmic rays may also be responsible for much of the
diffuse hard X-ray continuum emission. A detailed modeling of this process
will place interesting constraints on the largely-unknown population of 
low energy cosmic rays in the Galaxy.
\end{itemize}

The X-ray mapping of the GC region, together with observations 
of many other objects in the Galaxy, has allowed for detailed investigations
into various high energy
astrophysical phenomena and processes. But to gain galaxy-wind perspectives, 
we should resort to observations of nearby galaxies.

\section{Nearby Galaxies}

{\sl Chandra} has made numerous observations of nearby galaxies, from dwarf
irregulars to giant ellipticals. These observations
allow for examining the interplay between various galactic 
components as well as for studying populations of X-ray sources at the same
distances to individual galaxies and with minimum line-of-sight 
confusion. I comment here on several such topics.

\subsection{X-ray Source Populations}

Even with the {\sl Chandra} capability, we typically only detect X-ray 
sources with luminosities $\ga 10^{37} {\rm~erg~s^{-1}}$ in nearby galaxies 
outside the Local Group. Such X-ray sources are mostly X-ray binaries with 
neutron stars or black holes as the primaries. The secondaries of the binaries
are either high mass stars (HMXBs with lifetimes of $\sim 10^7$ yrs)
or low mass stars (LMXBs with timescale of $\sim 10^9$ yrs).
The relative populations of these two distinct types of X-ray binaries,
especially for the luminous ones,
naturally depend on galaxy types: starburst galaxies are prevailed by HMXBs
whereas early-type galaxies are dominated by LMXBs. What do we learn
from {\sl Chandra} observations?

\begin{itemize}
\item The high resolution observations of {\sl Chandra} have allowed
for the first time to reliably construct 
the X-ray source luminosity function, $N$ - log $S$, for individual galaxies.
It is found that the $N$ - log $S$ slope flattens
with increasing star formation rate of spiral galaxies (Kilgard et al.
2002; Swartz et al. 2002), which is 
consistent with the assumption that the more luminous HMXBs
are shorter-lived. Therefore, the slope, or the high luminosity cutoff,
of the luminosity function can be used as a measure of the massive star formation 
history of a galaxy (e.g., Wu 2001). However, the formation and evolution
of LMXBs are believed to be quite different from those of HMXBs, which
can seriously complicate the measurement.

\item Large populations of LMXBs have been revealed in elliptical 
galaxies (e.g., Sarazin, Irwin, \& Bregman 2000; Kraft et al. 2001; 
Angelini et al. 
2002). White, Sarazin, \& Kulkarni (2002)  find that the ratio of 
the global LMXB luminosity to the galactic optical luminosity 
is not correlated with optically-derived stellar age indicator and, 
instead, is strongly correlated with the specific globular cluster 
frequencies in elliptical galaxies. In an HST-imaged field of NGC 1399, 
for example,
Angelini et al. (2001) show that globular clusters account for $\sim 70\%$ of X-ray 
sources with luminosities $\ga 5 \times 10^{37} {\rm~ergs~s^{-1}}$.
This suggests that most LMXBs were formed in globular clusters. 

\item log $N$ - log $S$ relations for some elliptical galaxies show
a break approximately at the Eddington limit ($\sim 2 \times 10^{38} 
{\rm~ergs~s^{-1}}$) on the isotropic luminosity for accretion onto a 
1.4$M_\odot$ neutron star (Sarazin et al. 2000; Finoguenov \& Jones 2002).
This break, if confirmed, may be used as a distance indicator to nearby 
galaxies.

\end{itemize}

\subsection{Ultra-Luminous X-ray Sources}

A considerable number of X-ray sources outside galactic nuclei appear to have 
luminosities substantially greater than the 
Eddington limit for a few solar mass black hole, assuming that the emission is isotropic. 
Some of these sources have luminosities as
high as a few times $10^{40} {\rm~ergs~s^{-1}}$. X-ray sources of this type,
known for more than 20 years since the detections with {\sl Einstein}, are
often called ultra-luminous X-ray sources (ULXs) or super-Eddington
sources. The most commonly-adopted interpretation for ULXs 
assumes that they are binary systems containing the so-called intermediate-mass 
black holes ($M \sim 10^2-10^5 M_\odot$). Alternatively, the apparent high
X-ray flux may arise from highly beamed emission from accretion disks
around stellar mass black holes or neutron stars ---
the micro-quasar scenario (e.g., Markoff, Falcke, \& Fender 2001;
King 2002). Furthermore, the binaries could
also be truly super-Eddington, for example, obtainable from accretion disks with
radiation-driven inhomogeneities (Begelman 2002). Much of the current debate
is driven by new X-ray observations made with both {\sl Chandra}
and {\sl XMM-Newton}. Let us see how such scenarios may confront 
the observations:

\begin{itemize}
\item Aperiodic variability by a factor greater than 2 and on 
various timescale scales has been detected for several ULXs 
(e.g., Mukai et al. 2002; Pence et al. 2001; Strickland et al. 2001).  
Such strong variability
is expected for the relativistic beaming (jet) models.

\item Two ULXs have been shown to vary periodically (Bauer et al. 2001; 
Liu et al. 2002).
For example, a period of 7.5 hours is found for a ULX in the Circinus Galaxy.
This ULX has a mean X-ray luminosity of $3.4 \times 10^{39} {\rm~erg~s^{-1}}$
and an X-ray spectrum well fitted by a multicolor blackbody accretion 
disk model. These properties are consistent with an eclipsing $\ga 50
M_\odot$ black hole binary (Bauer et al. 2001).
The implied viewing angle, nearly edge-on, can probably be used to rule out
the relativistic beaming model for such ULXs, because a jet should
be oriented in the direction perpendicular to an accretion disk and hence 
should not be pointed to us.

\item Most of ULXs have remained remarkably steady throughout 
their observed history, which goes back as long as 20 years ago when 
{\sl Einstein} Observatory made the first detections (e.g., 
Swartz et al. 2000). This persistence of source fluxes poses difficulties for 
the jet models, especially ones that require large relativistic boosting of
X-ray fluxes.

\item Multiple observations show that some ULXs (e.g., Kubota et al. 2001; 
Liu, Bregman, \& Seitzer 2002) switched 
between high/soft and low/hard spectral states.
These state changes are characteristic of Galactic accreting black hole binaries.

\item An increasing number of ULXs have been identified in other wavelength bands
(e.g., Wu et al. 2002; Liu et al. 2002). In particular, many of ULXs
in elliptical galaxies are found within globular clusters.
Interestingly, black holes with masses $\sim 4 \times 10^3 M_\odot$ and 
$2 \times 10^4 M_\odot$ have been discovered recently at the centers of the 
globular clusters M15 in our Galaxy (Gerssen et al. 2002) and G3 in M31 
(Gebhardt, Rich, \& Ho 2002). Therefore, intermediate mass black holes
can originate in globular clusters, many of which may have been
tidally disrupted over the history of a galaxy. 

\item Many ULXs are located within shell-like optical nebulae
(e.g., see the Pakull's contribution in this volume).  The nebula
around the variable ULX in dwarf galaxy Holmberg II, in particular, 
shows strong 4686 \AA\ recombination radiation from HeII atoms. The ionization
of the atoms requires a soft X-ray luminosity that is consistent with the 
isotropic emission of the ULX.
\end{itemize}

Therefore, ULXs may represent a heterogeneous type of X-ray sources. 
It appears that accreting intermediate-mass black holes are a viable 
interpretation of most ULXs, although rapid
varying objects and/or transients may be more naturally explained by
the jet-induced beaming models or possibly by the super-Eddington scenarios.

\subsection{Hot Gas Outflows from Nuclear Starburst Galaxies}

The extreme behavior of the violent interplay between massive stars and
the ISM is the gas ejection from galactic disks. Naturally, most of the 
work on 
this topic has been concentrated on edge-on disk galaxies, which allow for
detection of extraplanar gas unambiguously. Vertical X-ray plumes are
detected along the minor axes of many nuclear starburst galaxies 
(e.g., NGC 253, Strickland et al. 2002; NGC 3079, Cecil et al. 2002;
NGC 4945, Weaver 2001; M82, Lehnert, Heckman, \& Weaver 1999, Fig. 4). These diffuse X-ray plumes are typically 
characterized by soft thermal spectra of plasma 
with steep temperature gradients in the disk/halo transition zones
close to the nuclear regions. Morphologically, the diffuse X-ray plumes
are often double horn-shaped and are well correlated with extraplanar H$\alpha$ 
features.

\begin{figure*}[!thb]
\unitlength1.0cm
    \begin{picture}(13,6.4) 
\put(0.,0){
          \begin{picture}(6.4,6.4)
	\end{picture}
	}
\put(6.6,0){
          \begin{picture}(6.4,6.4)
	\psfig{figure=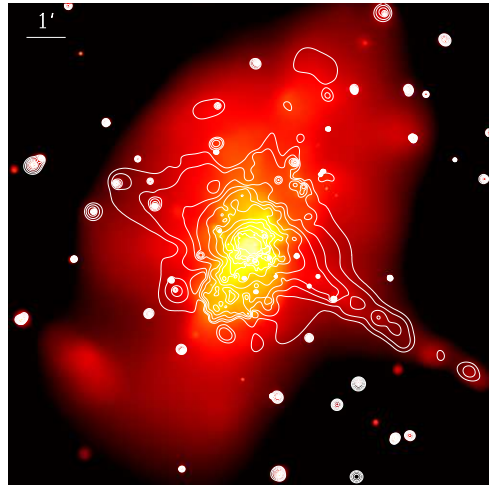,height=6.4cm,angle=90,clip=}
	\end{picture}
	}
    \end{picture}
\caption{M82 in H$\alpha$ (left panel; Hintzen et al. 1993) and in X-ray 
(right panel). The X-ray intensity image is in the 0.3-0.7 keV band, whereas
the contours are in 0.7-1.5 keV band. The linear extension toward the 
southwest (the CCD readout direction), nearly parallel to the disk of the galaxy, 
is partly an artifact. }
\end{figure*}

In comparison, outflows driven by AGNs are not necessarily along 
the minor axes of galaxies and are primarily in form of jets or 
well-collimated/centrally-filled lobes
 (e.g., NGC 1068; Young, Wilson \& Shopbell 2001). The X-ray emission from 
AGN-induced outflows is due to hot gas heated by the terminal shock
of the outflows (e.g., Terashima \& Wilson 2001).

Deep X-ray observations of nuclear starburst galaxies 
have further revealed large-scale, low surface 
brightness diffuse X-ray halos, which seem to be confined within faint 
limb-brightened H$\alpha$-emitting boundaries (e.g., NGC 253, 
Strickland et al. 2002; NGC 1569, Martin et al. 2002; M82, 
 Lehnert et al. 1999, Fig.\ 4).

Extraplanar diffuse X-ray emission has also been observed around 
relatively normal galaxies, in which star formation is not concentrated in 
nuclear regions. The best case for the presence of a diffuse hot gaseous halo is  
the edge-on Scd galaxy NGC 4631 (Wang et al. 2001). The diffuse X-ray 
emission is very soft and extends up to about 10 kpc away from the galactic 
plane. The overall X-ray emission morphologically resembles the 
well-known radio halo of the galaxy, 
indicating a possible link between outflows of hot gas and cosmic 
ray/magnetic field from the galactic disk. Substantial amounts of 
extraplanar diffuse X-ray emission have also been detected around another 
nearby Sc galaxy NGC 3556 (Wang 2002). Its radio and X-ray halos are also
morphologically similar. 

The most logical explanation for the characteristics of the hot gas
outflows and halos is a confined
galactic wind model (e.g., Strickland et al. 2002). A free-streaming wind
is not efficient in radiating X-rays, because it is cooled and rarefied
rapidly during the adiabatic expansion phase within or just outside the 
galactic disks or nuclear regions. Thus,  the X-ray emission 
that spatially correlates with H$\alpha$ features does not
directly arise from fast outflowing galactic winds. 
The emission may instead come from shocked and/or dynamically mixed gas 
around pre-existing clouds or at the interfaces between the winds and the
galactic gaseous disks. If sufficient 
amounts of cool gas are present in the halos, as indicated by the outer
H$\alpha$-emitting boundaries observed, the galactic wind may be 
thermalized again and may be mass-loaded by thermal conduction at the boundaries. 
The thermalized plasma can
thus be responsible for the large-scale, low surface brightness X-ray 
emission seen around M82, for example (Fig. 4). Of course, if the plasma
is hot enough and does not cool sufficiently fast, it may eventually escape 
from galaxies due to the buoyancy and/or to the ram-pressure of the IGM.

\subsection{Interplay of Galaxies with Their Environments}

\begin{figure*}[!htb]
 \begin{center}
\psfig{figure=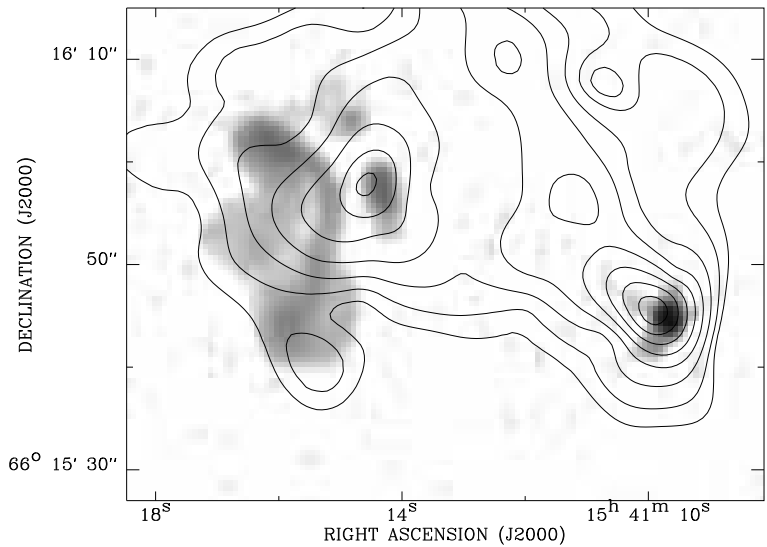,height=3.2truein,angle=0,clip=}
 \end{center}
\caption{The central region of the cluster 
Abell 2125 ($z = 0.25$) in radio (gray-scale) and X-ray (contours).
The X-ray intensity contours are at 14, 15, 18, 20, 24, 28, 33, and 39
$\times 10^{-3} {\rm~cts~s^{-1}~arcmin^{-2}}$ in the 0.5-4 keV band.
}
\end{figure*}

The large amounts
of gas in galactic halos, as inferred from the confinement of
galactic winds, may represent the accretion of the IGM onto the galaxies  
 in the group environment. 
The accretion is predicted by numerical simulations 
of the structure formation (e.g., Murali et al 2002) and is 
supported by lines of observational evidences (Burstein \& Blumenthal 2002). 
The accretion, likely consists of multiple gaseous phases,
may also occur around our Galaxy, as indicated by the presence of high 
velocity 
HI clouds of low metallicity and by recent detections of OVI, OVII, and OVIII 
absorption lines with {\sl FUSE} and {\sl Chandra} (e.g., Nicastro et al. 2002). 
The interaction between the galactic outflows and 
accretion inflows can then be used to probe their physical and chemical 
properties.

The effect of the intergalactic gas ram-pressure can be seen most vividly in 
the rich cluster environment. Fig.\ 5 demonstrates the use of X-ray emission
as a tracer for the effect (Wang et al. 2002, in preparation). 
The hot intracluster medium density clearly affects the morphology of 
extended radio lobes. Furthermore, an apparent X-ray-emitting wake,
trailing the lower right galaxy in the figure, is likely due to 
the ram-pressure 
stripping of galactic cool gas and possible subsequent mixing with the hot 
ambient medium. An ongoing multiwavelength investigation will help to 
quantify the interaction between these galaxies and the cluster environment.


I thank the symposium organizers for inviting me to give this talk and
my research collaborators for their contributions to the various projects mentioned
above.


\begin{references}

\reference Angelini, L., et al. 2002, ApJ, 568, L97
\reference Baganoff, F., et al. 2001, Nature, 413, 45
\reference Begelman, M. C. 2002, APJ, 568, L97
\reference Bauer, F. E., et al. 2001, AJ, 122, 182
\reference Burstein, D., \& Blumenthal, G. 2002, ApJ, 574, L17
\reference Cecil, G., Bland-Hawthorn, J., \& Veilleux, S. 2002, ApJ, 576, 745
\reference Chlebowski, T. 1989, ApJ, 342, 1091
\reference Dutra, C. M., \& Bica, E. 2000, A\&A, 359, L9
\reference Figer, D., et al. 1999, ApJ, 525, 750
\reference Finoguenov, A.,  \& Jones, C. 2002, ApJ, 574, 754
\reference Gebhardt, K., Rich, R. M.,  \& Ho, L. 2002, ApJ, 578, L41
\reference Gerssen, J., et al. 2002, AJ, in press
\reference Grindlay, J., Heinke, C., Edmonds, P., \& Murray, S. 2001, Science, 292, 2290
\reference Helfand, D. J. 1984, ASPP, 96, 913
\reference Hintzen, P., et al. 1993, In The Evolution of Galaxies and Their Environment, 38 
\reference Kilgard, R. E., et al. 2002, ApJ, 573, 138
\reference King, A. R. 2002, MNRAS, 335, L13
\reference Koyama, K., et al. 1996, PASJ, 48, 249
\reference Kraft, R. P., et al. 2001, ApJ, 560, 675
\reference Kubota, A., et al. 2001, ApJ, 547, L119
\reference Lang, C., et al. 2001, ApJL, 551, 143
\reference Lehnert, M. D., Heckman, T. M., \& Weaver, K. A. 1999, ApJ, 523, 575
\reference Liu, J. F., Bregman, J. N., \& Seitzer, P. 2002, ApJ, 580, L31
\reference Liu, J. F., Bregman, J. N., Irwin, J. A., \& Seitzer, P. 2002, ApJL, in press
\reference Lu, F., Wang, Q. D., \& Lang, C. C. 2002, ApJ, submitted
\reference Markoff, S., Falcke, H., \& Fender, R. 2001, A\&A, 372, L25
\reference Martin, C. L., Kobulnicky, H. A., \& Heckman, T. M. 2002, 574, 663
\reference Morris, M. 1994, in The Nuclei of Normal Galaxies, p185
\reference Mukai, K., et al. 2002, ApJ, in press (Astro-ph/0209166)
\reference Murali, C., et al. 2002, ApJ, 571, 1
\reference Nicastro, F., et al. 2002, Nature, submitted (astro-ph/0208012)
\reference Pence, W. D., et al. 2001, ApJ, 561, 189
\reference Pfahl, E., Rappaport, S., \& Podsiadlowski, P. 2002, ApJL, 571, 37
\reference Portegies Zwart, S. F., et al. 2002, ApJ, 565, 265
\reference Raga, A. C., et al. 2001, ApJL, 559, 33
\reference Sarazin, C. L., Irwin, J. A., \& Bregman, J. N. 2000,  ApJ, 544, L101
\reference Serabyn \& G\"usten 1987, A\&A, 184, 133
\reference Strickland, D. K., et al. 2001, ApJ, 560, 707
\reference Strickland, D. K., et al. 2002, ApJ, 568, 689
\reference Swartz, D. A., et al. 2002, ApJ, 574, 382
\reference Takagi, S., Murakami, H., \& Koyama, K. 2002, ApJ, 573, 275
\reference Terashima, Y., \& Wilson, A. S. 2001, Apj, 560, 139
\reference Young, A. J., Wilson, A. S., \& Shopbell, P. L. 2001, ApJ, 556, 6
\reference Valinia, A., et al.  2000, ApJ, 543, 733
\reference Wang, Q. D. 2001, in The New Vision of the X-ray Universe in the XMM-Newton and Chandra Era'', in press (astro-ph/0202317)
\reference Wang, Q. D., Gotthelf, E., \& Lang, C. 2002, Nature, 415, 148
\reference Wang Q. D., Lu, F., \& Lang, C. C. 2002, ApJ, in press (Dec. 20)
\reference Weaver, K. A. 2001, in The Central Kiloparsec of Starbursts and AGNs, APS
Conference Series 249, 389
\reference White III, R. E., Sarazin, C. L., \& Kulkarni, S. R. 2002, 571, L23
\reference Wijnands, R., \& Wang, Q. D., 2002, ApJL, 568, 93
\reference Wijnands, R., Miller, J. M., \& Wang, Q. D. 2002, ApJ, 579, 422
\reference Young, Y., Wilson, A. S., \& Shopbell, P. L. 2001, ApJ, 556, 6 
\reference Yusef-Zadeh, F., et al. 2002, ApJ, 570, 665
\reference White III, R. E., Sarazin, C. L., \& Kulkarni, S. R. 2002, ApJ, 571, L23
\reference Wu, H., et al. 2002, ApJ, 576, 738
\reference Wu, K. PASA, 18,443
\end{references}
\end{document}